\documentclass[journal=jpcafh,manuscript=article]{achemso}
\setkeys{acs}{doi = true}
\usepackage[version=3]{mhchem} 

\usepackage[separate-uncertainty=true,number-mode=math,unit-mode=text]{siunitx}
\usepackage{graphicx}
\usepackage{microtype}
\usepackage{amsfonts,amsmath,amssymb,upgreek}
\usepackage{tabularx}
\usepackage{bm}
\usepackage{wasysym} 
\usepackage{physics}
\usepackage{dcolumn}
\usepackage{ragged2e}
\usepackage{xcolor,soul}

\usepackage{hyperref}
\hypersetup{
	colorlinks=true, 
	linkcolor={red!65!black},
	citecolor={blue!75!black},
	urlcolor={blue!50!black}
}
\usepackage[capitalise]{cleveref} 
\creflabelformat{equation}{#2\textup{#1}#3} 

\usepackage[utf8]{inputenc}
\usepackage[T1]{fontenc}
\usepackage{mathptmx}
\usepackage{etoolbox}
\usepackage[relative]{overpic} 

\newcommand{\doi}[1]{\href{http://dx.doi.org/#1}{\nolinkurl{#1}}}
\makeatletter
\newcommand\thefontsize{The current font size is: \f@size pt}
\makeatother

\author{Stach E.\,J. Kuijpers}
\author{Panagiotis Kalaitzis}
\author{Evangelia Sakkoula}
\author{Sebastiaan Y.\,T. van de Meerakker}
\affiliation{Radboud University Nijmegen, Institute for Molecules and Materials, Heijendaalseweg 135, 6525 AJ Nijmegen, The Netherlands}
\author{Timothy P. Softley}
\affiliation{School of Chemistry, University of Birmingham, Edgbaston, Birmingham B15 2TT, United Kingdom}
\author{David H. Parker}
\email{parker@science.ru.nl}
\affiliation{Radboud University Nijmegen, Institute for Molecules and Materials, Heijendaalseweg 135,
	6525 AJ Nijmegen, The Netherlands}

\title{Sensitive Low-Recoil VUV 1+1$'$ REMPI Detection of ND$_3$}


\begin{document}


%
%
%



\begin{abstract}
In  molecular beam scattering experiments, an important technique for measuring product energy and angular distributions is velocity map imaging following photoionization of one or more scattered species. For studies with cold molecular beams, the ultimate resolution of such a study is often limited by the product detection process. When state-selective ionization detection is used, excess energy from the ionization step can transfer to kinetic energy in the target molecular ion--electron pair, resulting in measurable cation recoil.  With state-of-the-art molecular beam technology, velocity spreads as small as a few \si{m/s} are possible, thus a suitable product detection scheme must be not only highly sensitive, state-selective, and background-free, it must also produce significantly less cation recoil than the velocity spread of the molecular beams undergoing cold collisions. To date this has only been possible with the NO molecule, and our goal here is to extend this minimal-recoil capability to the fully deuterated ammonia molecule, ND$_3$. In this article a resonance enhanced multi photon ionization (REMPI) detection scheme for ND$_3$ is presented that imparts sufficiently low recoil energy to the ions, allowing, for the first time, high-resolution imaging of ND$_3$ collision products in cold molecule scattering experiments with HD. The excitation step of the 1+1$'$ REMPI scheme requires vacuum ultra-violet (VUV) photons of \SI{\sim160}{nm}, which are generated through four-wave-mixing in Xe. We varied the wavelength of the second, ionization step between \num{434} and \SI{458}{nm}, exciting ND$_3$ to a wide range of autoionizing neutral states. By velocity mapping the photoelectrons resulting from the detection scheme, it was possible to fully chart the ion recoil across this range with vibrational resolution for the final ionic states. Additionally, rotational resolution in the photoionization dynamics was achieved for selected excitation energies near one of the vibrational thresholds. Many of the peaks in the spectrum of autoionizing Rydberg states are assigned to specific Rydberg series using a simple Rydberg formula model. 
\end{abstract}

\newpage
\section{Introduction}

State-specific and efficient detection is a highly important ingredient in many molecular dynamics experiments. The \textit{de facto} way to detect ammonia in gas-phase experiments is by 2+1 REMPI through the $B\,^1E''$ state (the $3p_{x\text{-}y}$ Rydberg state) using ultraviolet (UV) light around \SI{320}{nm}, followed by time-of-flight detection of the formed ions. This fully state-selective ionization scheme has been used in spectroscopy to study NH$_3$ and ND$_3$ in great detail, both for the $B$ state \cite{Ashfold:CPL138:201,Ashfold:JCP89:1754,Langford1998,Bentley2000}, as well as several higher lying Rydberg states \cite{Ashfold1986,Ashfold1988,Ashfold1998,Langford1998}. The $A\,^1A_2''$ state of ammonia (the $3s$ Rydberg state) is highly dissociative \cite{Baronavski1995}, making it unsuitable as an intermediate. Via the $B$ state, it is possible to detect the inversion doublet components of ground state ND$_3$ separately, despite their small splitting of \SI{0.053}{\per\cm}, since each connects to different vibrational levels of the excited state by virtue of the parity selection rule. The Rydberg states of ammonia are all of planar equilibrium geometry, following promotion of a ground state lone pair electron. Because of this change in geometry compared to the pyramidal ground state, long vibrational progressions in the inversion mode $\nu_2$ are observed in transitions from the ground state. For the $B(\nu'_2) \leftarrow X(\nu_2)$ transition in ND$_3$, the $2_0^8$ band ($\nu'_2=8 \leftarrow \nu_2=0$) has the largest absorption cross section \cite{Cheng2006}.

Beyond direct spectroscopic studies, efficient detection of nascent ammonia through the $B$ state has been a key tool for experimentalists, making ammonia the molecule of choice for many proof-of-principle experiments on the manipulation of neutral polar molecules. This detection scheme was used to demonstrate a combination of Stark deceleration and electrostatic trapping \cite{Bethlem:Nature406:491,Bethlem:PRA65:053416} and AC trapping of high-field-seeking states \cite{Veldhoven:PRL94:083001,Schnell:JPCA111:7411} as well as the operation of a molecular buncher \cite{Crompvoets:PRL89:093004}, mirror \cite{Schulz:PRL93:020406}, storage ring \cite{Crompvoets:Nat411:174}, synchrotron \cite{Heiner:NatPhys3:115,Zieger2010}, beamsplitter \cite{Deng2011,Gordon2017:3D} and fountain \cite{Cheng:PRL117:253201} for the first time.
The excellent sensitivity of the 2+1 REMPI scheme made it possible to record high resolution microwave spectra on the $N_K=1_1$ inversion doublet of $^{14}$ND$_3$ and $^{15}$ND$_3$ \cite{Veldhoven:PRA66:032501} and record infrared action spectra of the ammonia-water dimer \cite{Mollner2009}. 
The scheme was also used to probe the orientation of buffer-gas-cooled ammonia molecules emerging from a quadrupole guide velocity selector \cite{Steer2017} and study collisions between cotrapped ND$_3$ molecules and laser cooled Rb atoms \cite{Parazzoli:PRL106:193201,Fitch2020}.
Finally, ionization through the $B$ state has been applied together with VMI in crossed-beam collision experiments to detect the angular distribution of ND$_3$ molecules after scattering with He \cite{Tkac2014} and H$_2$ \cite{Tkac2015,Gao2019}.

However, there are two drawbacks to using the 2+1 REMPI scheme for velocity-resolved scattering experiments using VMI. First, the two-photon step requires a tightly focused laser, such that only a small part of the active volume of the experiment is probed. More importantly, since the total photon energy far exceeds the ionization threshold, a recoil is imparted to the formed ion, obscuring the neutral's velocity distribution of interest and spoiling the resolution of experimental images. For ionization of ND$_3$ via $B(\nu'_2=5)$, this recoil is \SI{\sim17}{m/s}, which has prevented scattering experiments with Stark-decelerated ND$_3$ packets from reaching their ultimate resolution \cite{Gao2019}. In principle, coincidence detection of the ion and photoelectron retrieves sufficient information to reconstruct the molecules' velocity prior to ionization, but this method requires a more complicated VMI detector and is only easy to implement when a single ion-electron pair is detected per laser shot. The availability of a state-selective, efficient \textit{and} recoil-free detection scheme for ND$_3$ would solve this resolution problem at the source.

In a 1+1$'$ REMPI scheme, vacuum ultra-violet (VUV) photons with a \SI{160}{nm} wavelength can be used to directly excite the $B \leftarrow X$ transition reaching the $\nu'_2=5$ or $6$ levels of the $B$ state. Ionization then takes place by absorbing a second photon of appropriate wavelength to excite into the region of the ionization threshold with little excess energy. A 2+1$'$ REMPI scheme could achieve the same low recoil, but would suffer from competition with the high-recoil 2+1 pathway. The required wavelength of \SI{160}{nm} can be generated through four-wave mixing in a nonlinear medium, usually a rare gas. More specifically, difference frequency mixing (DFM) can be applied \cite{Hilbig1983,Miyazaki1989}, exploiting optical transitions of the medium for a higher conversion efficiency. Previous implementations at a VUV wavelength of \SI{\sim160}{nm} include probing the dynamics of ammonia clusters at femtosecond time scales (DFM in Ar) \cite{Freudenberg1997}, laser-induced fluorescence detection of C and CO (DFM in Xe) \cite{Chastaing2000,Carty2004} and REMPI detection of NH$_3$, C and CO (DFM in Kr or Xe) \cite{Perez:thesis:2014,Plomp2021,Plomp:thesis:2023,Suits2014,Sun2017}. The VUV detection of C atoms was applied successfully to image collision products in crossed-beam scattering with He and H$_2$ at high resolution. Additionally, using VUV light generated by sum frequency mixing, one-photon transitions to the higher vibrational levels of the $B$ state of ND$_3$ and NH$_3$ have been observed \cite{Li1994,Li1995}. Although DFM is an inefficient process yielding small amounts of VUV light per shot, the transition probability to absorb a single VUV photon and reach the $B$ state of NH$_3$ or ND$_3$ is high. Overall, 1+1$'$ signal levels can be comparable to the 2+1 REMPI scheme \cite{Perez:thesis:2014}, while a larger active volume of the experiment can be probed. The vibrational one-photon VUV absorption cross sections of the $B \leftarrow X$ transition in both ND$_3$ and NH$_3$ are well known from experiments with synchrotron radiation \cite{Cheng2006,Pratt2023}, while the rotational spectroscopy is fully explored by two-photon experiments, as described above \cite{Ashfold:JCP89:1754}.

In contrast to the well-understood $B \leftarrow X$ transition, the $X^+\leftarrow B$ ionization step is both more complicated and less well studied, particularly for ND$_3$, as the ionic ground state rovibrational manifold $X^+(\nu^+,N^+_{K^+})$ has not been fully mapped, especially for levels with $\nu_2^+>2$. Additionally, each state in this manifold has several Rydberg series of neutral states converging toward it, differing by the quantum numbers of the outer-electron orbital angular momentum and the total angular momentum (the vector sum of the Rydberg-electron and ion-core angular momenta). Rydberg states lying above the ionization threshold may autoionize efficiently, leading to a complex photoionization (PI) spectrum with features from many overlapping Rydberg series (see below). The NH$_3^+$ and ND$_3^+$ vibrational states have been probed through one-photon ionization and UV photoelectron (PE) spectroscopy, although the origin band of the vibrational ladder for ND$_3$ could not be assigned definitively due to the presence of the $2_1^0$ hot band \cite{Weiss1970,Potts1972,Rabalais1973,Locht1991,Locht1998}. Since the ion has a planar geometry, a long progression in $\nu^+_2$ was observed. Also, activity in the $\nu^+_1$ and $\nu^+_4$ vibrational modes was observed. The relative energies of the vibrational states $X^+(\nu^+)$ have been calculated theoretically \cite{Botschwina1988,Leonard2002}. For NH$_3$, several studies on threshold ionization have been performed using a 2+1$'$ REMPI scheme. zero kinetic energy (ZEKE) photoelectron spectra were recorded in the $\nu_2^+=1,2$ region using a delayed extraction field \cite{Habenicht:JCP95:4809}. Pulsed-field ionization was used to detect Rydberg states just below the ionization potential, yielding mass-analyzed threshold ionization (MATI) \cite{Dickinson1997,Dickinson:JPCA105:5590} ion spectra. Both ZEKE and MATI spectra could be reproduced using multichannel quantum defect theory (MQDT) simulations, which take into account the various ionization thresholds (ionic energy levels) and the Rydberg series (including their mutual perturbations) converging to each of them. Rotationally resolved PE spectra were recorded using electron VMI (eVMI) in the $X^+(\nu^+_2=4)$ region of NH$_3$ \cite{Hockett2010}. Finally, one-photon PI spectra of both NH$_3$ and ND$_3$ as well as the two mixed isotopomers were recorded with \SI{0.008}{\per\cm} resolution in the vicinity of the lowest ionization thresholds, supplemented by several PFI spectra for NH$_3$ \cite{Seiler2003} and the mixed isotopomers \cite{Hollenstein2007}. These measurements were later extended for ND$_3$ to include the $\nu^+_2=2$ threshold region, along with an MQDT simulation reproducing \SI{80}{\percent} of the spectral lines \cite{Duggan2010}. The adiabatic ionization energy connecting the ground states of ND$_3$ and ND$_3^+$ was reported to be \SI{82261.7}{\per\cm}, establishing the correct vibrational level labeling of the photoelectron spectrum bands.

We report here a state-selective, low-recoil 1+1$'$ REMPI scheme for ND$_3$ at $160+448$~\si{nm}, using VUV photons generated by DFM to excite to the intermediate $B(\nu'_2=5,6)$ states, which are perturbation-free and at convenient wavelengths near the maximum of the Franck-Condon envelope. PI spectra covering the $\Delta\nu_2=\nu_2^+-\nu'_2=-2,-1,0$ regions are recorded, resolving many autoionizing Rydberg states and showing a propensity for excitation to Rydberg states with $\Delta\nu_2=0$. eVMI is used to record wavelength dependent PE spectra with vibrational resolution, characterizing the photoionization dynamics and resulting ion recoil. While small changes in the vibrational quantum numbers are to be expected upon autoionization, such behavior is not a given in this energy range of a polyatomic. We address Rydberg states that can autoionize on energetic grounds to more than 20 open vibrational channels (across all modes), but observe a predominance of low-energy electrons, which is critical to achieving our objective of low ion-recoil. Rotationally resolved PE kinetic energy two-dimensional (2D) spectra are recorded to study the ionic rotational state energies of $\nu_2^+=5$ and the relative propensities for their formation in the autoionization process. Many of the lines in the experimental photoionization spectra are assigned to specific Rydberg series using parameters derived from an MQDT calculation. Finally, the efficacy of this 1+1$'$ REMPI scheme for scattering experiments is demonstrated by recording a low-energy scattering image of ND$_3$ colliding inelastically with HD.

\section{Experimental Section}
\label{sec:eVMI_methods}
PI spectra and scattering images were recorded with the crossed molecular beam scattering apparatus described in Ref.~\citenum{Jongh2020}. A \SI{2.6}{m} long Stark decelerator was used to prepare a packet of ND$_3$ with population predominantly in the $N_K^p=1_1^-$ level. In order to observe transitions starting from the $1_1^+$ level, the packet of ND$_3$ molecules emerging from the decelerator could be converted to ($1_1^+$) via a microwave-induced transition at \SI{1.6}{\GHz}. The microwaves were generated by a $\lambda/4$ monopole antenna positioned inside the vacuum setup, connected to a signal generator (Rhode and Schwarz SMA100B). Using single-frequency pulses, the population transfer efficiency of this transition was limited to \SI{50}{\percent} \cite{Herbers2022}. Next, ND$_3$ was ionized by the 1+1$'$ REMPI scheme and accelerated toward the detector by a VMI lens. The extraction field in this setup used for PI spectra was \SI{20}{V/cm} based on the applied voltages and SIMION simulations of the lens geometry as in Ref.~\citenum{Plomp2020}.

ND$_3$ ($1_1^-)$ could be detected via the $B(\nu'_2=6,N'_{K'}=2_2) \leftarrow X(\nu_2=0,N_K^p=1_1^-)$ transition at \SI{63754.44}{\per\cm} (\SI{156.85}{nm}). Similarly, $1_1^+$ was detected via the $B(\nu'_2=5,N'_{K'}=2_2) \leftarrow X(\nu_2=0,N_K^p=1_1^+)$ transition at \SI{62995.09}{\per\cm} (\SI{158.74}{nm}). In both cases, the transition overlaps with a $1_1 \leftarrow 0_0^\pm$ transition. However, since the $0_0^\pm$ states do not exhibit a strong Stark effect, they are eliminated from the experiment by our state selectors. From the $B$ state, ionization could proceed after absorbing a second, blue photon of \SI{448}{nm}. This wavelength was selected to drive transitions that were diagonal in $\nu_2$ between the intermediate state and the final states (i.e. close to the ionization thresholds for $\nu_2^+=6$ or $5$ in the ion); such transitions were expected to be most intense because of the closely similar geometry of the $B$ state and the higher Rydberg states.

VUV radiation was generated through DFM by focusing two lasers into a gas cell filled with \SI{25}{m\bar} of Xe. The first laser was near-resonant with a two-photon transition in Xe at a wavelength of $\lambda_\text{R}\approx\SI{250}{nm}$, and pulse energy of $P_\text{R}=\SI{5}{mJ}$. The second laser was used to tune the wavelength of the generated VUV radiation. It had a wavelength of $\lambda_\text{T}\approx\SI{610}{nm}$, and a pulse energy of $P_\text{T}=\SI{10}{mJ}$. The beam subsequently passed through a sealed box where the VUV radiation was separated from the incoming colors by a CaF$_2$ Pellin-Broca prism. Nitrogen overpressure was applied to the box to prevent absorption of the VUV photons by oxygen.

Next, the VUV beam was collimated by a MgF$_2$ lens and overlapped with the ionizing, blue laser before entering the vacuum setup perpendicular to the ND$_3$ beam. The blue laser had a pulse energy of $P_\text{blue}=\SI{4}{mJ}$. The required wavelengths are summarized in \cref{tab:laser_wavelengths_VUV}. The two dye lasers used for DFM in Xe were pumped by the doubled and tripled output of a single YAG laser. The third dye laser used to produce the \SI{448}{nm} light was either pumped by the tripled output of the same source, or a separate YAG laser whose relative timing was controlled by a digital delay-pulse generator.

\begin{table}[h!]
	\caption{ Laser Wavelengths Used to Detect ND$_3$($1_1^\pm$)\textsuperscript{\emph{a}} }
	\label{tab:laser_wavelengths_VUV}
	\begin{tabular}{*6{c}}
		\hline \hline
		\noalign{\vspace{0.1cm}}
		$X(N_K^p)$ & $B(\nu'_2)$ & $\lambda_\text{R}$ (\si{nm}) & $\lambda_\text{T}$ (\si{nm}) & $\lambda_\text{VUV}$ (\si{nm}) & $\lambda_\text{blue}$ (\si{nm}) \\
		\hline
		\noalign{\vspace{0.1cm}}
		$1_1^+$ & 5 & 252.486 & 616.623 & 158.743 & 448.242 \\
		$1_1^-$ & 6 & 249.629 & 611.078 & 156.852 & 448.091 \\
		\hline \hline
	\end{tabular}
	\\ \vspace{0.2cm}
	\small{
		\textsuperscript{\emph{a}} Designating the energy of the different wavelengths $\lambda_i$ as $\nu_i$, DFM in Xe yielded VUV photons with $\nu_\text{VUV}=2\nu_\text{R}-\nu_\text{T}$. $\lambda_\text{R}$ is near resonant with a two-photon transition of Xe, while $\lambda_\text{T}$ can be tuned. $\lambda_\text{blue}$ ionizes the molecule from the $B$ state.
	}
\end{table}

Photoelectron imaging experiments were performed using a separate molecular beam machine. \SI{2}{\percent} ND$_3$ in Ar was supersonically expanded from a Nijmegen Pulsed Valve \cite{Yan:RSI84:023102} at a backing pressure of \SI{2}{\bar} and chamber pressure of \SI{2e-6}{m\bar}. The resulting beam entered the main vacuum chamber through a \diameter\SI{3}{mm} skimmer placed \SI{75}{mm} downstream from the nozzle. Next, ND$_3$ ($1_1^-$) was selected and focused toward the detection region by two \SI{10}{cm} long hexapoles with \diameter\SI{4}{mm} rods at a voltage of $\pm\SI{2}{kV}$. The two segments were placed \SI{7}{mm} apart, with a translatable \diameter\SI{3}{mm} beam stop in between. Other species were either blocked by this beam stop or focused away from the detection region. Next, the ND$_3$ molecules entered a conventional VMI detector, collinear with the beam axis, through a \diameter\SI{2}{mm} hole in the repeller. The center of the VMI detector was placed \SI{47}{cm} downstream from the skimmer, \SI{20}{cm} from the end of the hexapole. In between the VMI plates, ND$_3$ ($1_1^+$) could again be produced by a microwave-induced transition. The ND$_3$ was ionized using the 1+1$'$ REMPI scheme, generating photoelectrons that were mapped onto a gated detector, \SI{30}{cm} downstream. The two laser beams copropagate, perpendicular to the molecular beam. To reduce the amount of background electrons generated by stray VUV light, the repeller and extractor plates were placed \SI{20}{mm} apart, while \diameter\SI{3}{mm} pinholes were placed in vacuum before and after the VMI lens.

To record electron images, two combinations of VMI voltages for the repeller and extractor, $U_\text{R}$, $U_\text{E}$ were applied. The second extractor was grounded. At $U_\text{R}=\SI{-550}{V}$, $U_\text{E}=\SI{-335}{V}$, images with vibrational resolution were recorded at high throughput by integrating the signal appearing on the camera for 10-100 frames. With this rapid acquisition rate, 2D electron spectra could be recorded (electron kinetic energy versus ionization laser wavelength). The second REMPI wavelength was scanned in $1$ to \SI{1.5}{nm} intervals, with a \SI{0.003}{nm} step size. The initial and final wavelengths of each interval were calibrated with a wavemeter (HighFinesse, WS6-600), such that the wavelength of intermediate data points could be corrected by linear interpolation. At $U_\text{R}=\SI{-150}{V}$, $U_\text{E}=\SI{-90}{V}$, the center part of the electron images could be recorded with rotational resolution by additionally operating at sufficiently reduced laser power to detect only a few electrons per shot. Using event-counting and centroiding \cite{Chang1998,Li:RSI76:063106}, a high-resolution image could be accumulated over several hours. At these settings, the extraction field at the point of ionization was \SI{39}{V/cm} based on SIMION simulations.

\begin{figure}[b!]
	\centering
	\includegraphics[scale=1.2]{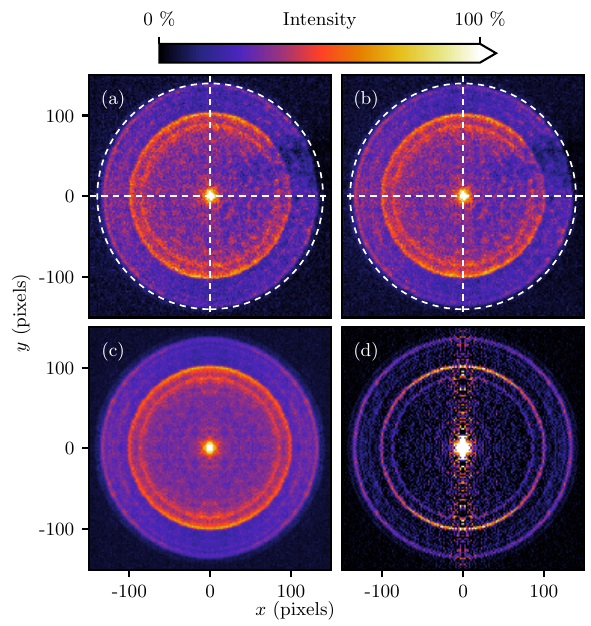}\hspace{1.5cm}
	\caption{Photoelectron image analysis. In four steps, the image is (a) centered, (b) circularized, (c) symmetrized and (d) Abel inverted. This image was recorded with the excitation laser set to the $B(\nu'_2=6,N'_{K'}=2_2) \leftarrow X(\nu_2=0,N_K^p=1_1^-)$ transition of ND$_3$, and the ionization laser set to $\lambda_\text{blue}=448.65$~\si{nm}. The color scale is identical for (a-c), but was rescaled for (d). White dashed lines in (a,b) guide the eye to the image center and a perfect circle. The upper-right quadrant of (b) was discarded to generate (c,d).
	}
	\label{fig:image_analysis}
\end{figure}

\begin{figure}[bt!]
	\centering
	\includegraphics[scale=1.2]{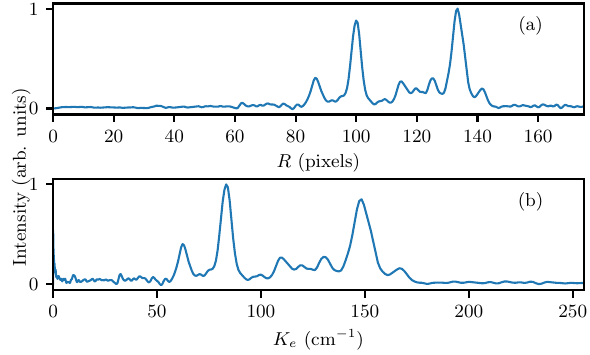}
	\caption{Distributions extracted from the photoelectron image of \cref{fig:image_analysis}. (a) radial intensity distribution $I(R)$, obtained by angular integration of the 3D Newton sphere that has \cref{fig:image_analysis}(d) as its central slice. (b) electron-kinetic energy distribution $I(K_e)$ (PE spectrum) obtained from $I(R)$. Peaks are assigned in \cref{fig:Ke_rotational}.  The upper-right quadrant of \cref{fig:image_analysis}(b) was included to generate these curves.
	}
	\label{fig:radial_analysis}
\end{figure}

Electron images were analyzed in four steps using the PyAbel \cite{Hickstein2019} package, as illustrated for a typical image in \cref{fig:image_analysis}(a-d). The images were centered manually (a) and circularized by stretching them by \SI{3}{\percent} along the $x$-axis (b). Circularization was only required for the high-resolution images recorded with a repeller voltage of \SI{-150}{V}, presumably to correct distortions introduced by stray fields, misalignment of the camera, or a lens distortion. A dead-spot is visible in the upper-right quadrant of the image, revealing a damaged area of the detector. Next, the images were symmetrized along both axes (c) and Abel inverted using the onion peeling method \cite{Bordas1996,Rallis2014} (d), which reconstructs the electron distribution along the center slice of the Newton sphere. The Abel inversion additionally returned the radial intensity distribution $I(R)$, which is shown in \cref{fig:radial_analysis}(a). The kinetic energy distribution of the electrons $I(K_e)$ follows since $K_e=f_\text{VMI}\cdot R^2$ for VMI detectors \cite{Eppink:RSI68:3477}, where $f_\text{VMI}$ is a scalar. $f_\text{VMI}$ was calibrated by comparing several images at different ionization wavelengths, and aligning the features in the kinetic energy distributions using $f_\text{VMI}$ as a free parameter. We found $f_\text{VMI}=1/40$~\si{cm^{-1}/pixel^2} for $U_\text{R}={\SI{-550}{V}}$ and $f_\text{VMI}=1/120$~\si{cm^{-1}/pixel^2} for $U_\text{R}={\SI{-150}{V}}$. \Cref{fig:radial_analysis}(b) shows the resulting electron-kinetic energy distribution. The quadrant containing the dead-spot could be discarded entirely during analysis. However, we found that the radial distribution of the electrons, which is of prime interest, is hardly changed by including this quadrant anyway. Hence, we chose to include it to compute the radial profiles, at the benefit of a slightly better signal-to-noise.

\section{Results and Discussion}
\subsection{Photoionization Spectra}
\label{subsec:PI_spectra}
\begin{figure}[b!]
	\centering
	\includegraphics[scale=1.2]{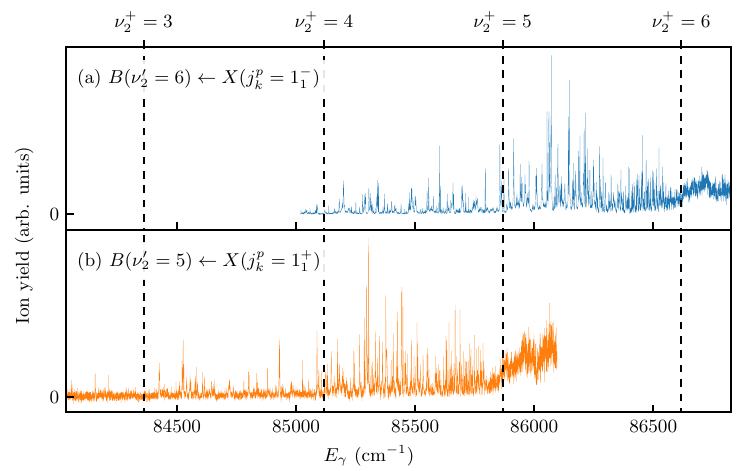}
	\caption{PI spectra of ND$_3$ recorded by scanning the ionization laser, following VUV excitation to the $B(\nu'_2=6,N'_{K'}=2_2)$ (a) and $B(\nu'_2=5,N'_{K'}=2_2)$ (b) state. $E_\gamma$ is the summed photon energy. Vertical dashed lines indicate the vibrational thresholds from Ref.~\citenum{Locht1998}, shifted by one quantum \cite{Duggan2010}.
	}
	\label{fig:PI_spectra}
\end{figure}

\Cref{fig:PI_spectra} shows the PI spectra recorded by fixing the VUV excitation laser as in \cref{tab:laser_wavelengths_VUV}, scanning the ionization laser in the $433-\SI{475}{nm}$ range and recording the photoion signal as a function of the summed energy of both lasers, $E_\gamma$. Vibrational thresholds reported in Ref.~\citenum{Locht1998} are shown as vertical dashed lines. The assignment of these thresholds shown in the figure has been shifted by one quantum to match the first three vibrational thresholds with those determined in Ref.~\citenum{Duggan2010}, and the positions moved by another \SI{8.3}{\per\cm} to account for the energy of the initial $X(1_1)$ state above the vibration-rotation ground state of ND$_3$. Even though ionization takes place from a single quantum state, $B(\nu'_2=6(5),N'_{K'}=2_2)$ for \cref{fig:PI_spectra}(a(b)), the PI spectra are crowded and highly structured due to the numerous series of autoionizing Rydberg states in this region converging to thresholds for different vibration-rotation states of ND$_3^+$. Apart from the vibrational separation the $\nu'_2=5$ and $\nu'_2=6$ spectra are nearly identical in their detailed structure. Overlapping them so as to match the majority of features requires an energy shift of \SI{766.7}{\per\cm}, which is an accurate measure for the vibrational spacing between $\nu_2^+=5,6$ assuming the rotational constants of these two ionic states are similar. As discussed below the rotational selection/propensity rules applicable to these spectra are identical, hence the observed close similarity of structure in the two spectra. This vibrational interval exceeds the $\nu_2'=5$ to $6$ vibrational spacing in the $B$ ($3p$) state by \SI{7}{\per\cm} \cite{Ashfold:JCP89:1754}. For Rydberg states with increasingly high principal quantum number, their rotational and vibrational parameters are expected to converge toward those of the ionic ground state. Indeed, the observed 5-6 vibrational spacing for the ionic thresholds does match, for example, that reported for the $F'$ ($5p$) state \cite{Langford1998}.

Since the $B$ and $X^+$ states are both planar and have similar vibrational frequencies, a propensity for diagonal ($\Delta\nu_2=0$) direct photoionization is expected, and a strong continuum arises in the observed spectra above the diagonal ionization threshold. However, as the excitation wavenumber passes below the $\Delta\nu_2=0$ threshold, a continuity of excitation probability is observed to $\Delta\nu_2=0$ Rydberg series appearing in the $\Delta\nu_2=-1$ photoionization region. These result from diagonal excitation to high-lying Rydberg states, which subsequently autoionize efficiently. Moreover, these Rydberg states couple most strongly to the $\Delta\nu_2=-1$ photoionization continua, and the spectra promptly become more sparse below the $\Delta\nu_2=-1$ threshold. These spectra can largely be modeled and assigned, as will be discussed below. First, photoelectron spectra are presented to characterize the recoil and autoionization dynamics in the region between $\Delta\nu_2=-1,0$, and probe the rotational constants of the $X^+(\nu_2^+=5)$ state. The assignment of the photoelectron spectra also gives more information on the rotational state assignment of the highly-lying Rydberg state resonances.

\subsection{Wavelength-Dependent Autoionization Dynamics}
\label{subsec:2D_spectra}
By recording photoelectron images while scanning the wavelength of the ionization laser, 2D PE spectra that reveal information on the autoionization dynamics of ND$_3$ could be gathered. We fixed the excitation laser to the $B(\nu'_2=6,N'_{K'}=2_2) \leftarrow X(\nu_2=0,N_K^p=1_1^-)$ transition and scanned the ionization laser to reach photon energies in the region around the $\nu_2^+=5,6$ thresholds. After extracting the radial distribution from each of the more than 8000 images, a 2D map of the wavelength-dependent electron kinetic energy distribution could be created.

\begin{figure}
	\centering
	\includegraphics[scale=1.2]{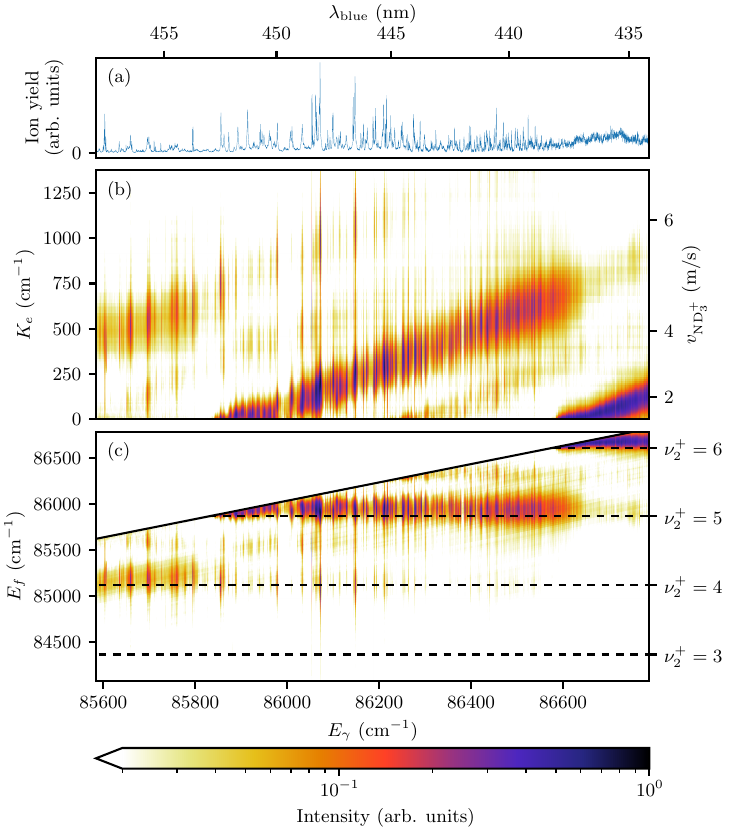}
	\caption{2D spectra of ND$_3$ obtained by scanning the ionization wavelength $\lambda_\text{blue}$ while velocity mapping the photoelectrons. The excitation wavelength is fixed on $B(\nu'_2=6,N'_{K'}=2_2) \leftarrow X(\nu_2=0,N_K^p=1_1^-)$, yielding a total photon energy of $E_\gamma$.
		(a) PI spectrum as in \cref{fig:PI_spectra}(a).
		(b) Relative intensity as a function of $E_\gamma$ and the electron kinetic energy, $K_e$. The secondary axes show the corresponding ion recoil, $\smash{v_\text{ND$_3^+$}}$ and $\lambda_\text{blue}$.
		(c) Relative intensity as a function of $E_\gamma$ and the final energy of the ion $E_f$ (relative to $\smash{X(N_K^p=1_1^-)}$). Horizontal black dashed lines mark the ionic vibrational thresholds as in \cref{fig:PI_spectra}. Relative intensity of (a,b) is expressed on a logarithmic color scale.
	}
	\label{fig:2D_spectra}
\end{figure}

\Cref{fig:2D_spectra}(b) shows this map for the $1_1^-$ state. With our interest in a low-recoil detection scheme in mind, this figure can be used to pick a wavelength with a strong intensity but low ion recoil $v_\mathrm{ND_3^+}$, as is shown along the right axis. Apart from the sharp resonances as a function of $E_\gamma$ in \cref{fig:2D_spectra}(b), which match the PI spectrum (\cref{fig:2D_spectra}(a)), five diagonal signal traces are visible. Each corresponds to a vibrational state of ND$_3^+$ and intersects the $x$-axis at its threshold energy. Some traces appear flattened for $K_e<\SI{100}{\per\cm}$ due to the finite imaging resolution, which sets a lower limit for the observed electron kinetic energy of ca. \SI{100}{\per\cm} at this high acquisition rate. Increasing $E_\gamma$ above threshold leaves a surplus of energy, which is converted into kinetic energy upon ionization. The final energy $E_f$ of the ion relative to the initial neutral $X(1_1^-)$ state is given by

\begin{equation}
	\label{eq:Ef}
	E_{f'} = E_\gamma - Ke \cdot (1+\frac{m_e}{m_\mathrm{ND_3}}).
\end{equation}

\noindent However, the presence of an electric field $F$ distorts the long-range Coulomb potential, lowering the ionization energy by $(\SI{6.1}{\per\cm})\cdot\sqrt{F/(\si{V/cm})}$ \cite{Merkt1998}. Thus, the VMI extraction field (see Experimental Section), which is not pulsed in our setup, induces a shift, which we correct by defining the field-free energy of the ionic state as

\begin{equation}
	\label{eq:Ef0}
	E_f = E_\gamma - Ke \cdot (1+\frac{m_e}{m_\mathrm{ND_3}}) + (\SI{6.1}{\per\cm})\cdot\sqrt{\frac{F}{\si{V/cm}}}.
\end{equation}

\begin{figure}
	\centering
	\includegraphics[scale=1.2]{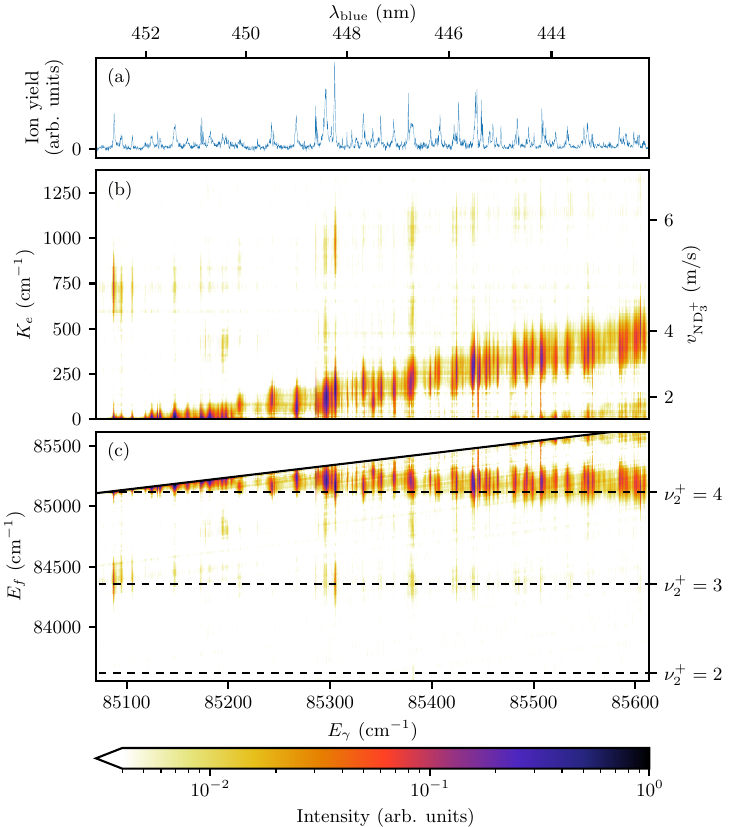}
	\caption{2D spectra of ND$_3$ obtained by scanning the ionization wavelength $\lambda_\text{blue}$ while velocity mapping the photoelectrons. The excitation wavelength is fixed on $B(\nu'_2=5,N'_{K'}=2_2) \leftarrow X(\nu_2=0,N_K^p=1_1^+)$, yielding a total photon energy of $E_\gamma$.
		(a) PI spectrum as in \cref{fig:PI_spectra}(b).
		(b) Relative intensity as a function of $E_\gamma$ and the electron kinetic energy, $K_e$. The secondary axes show the corresponding ion recoil, $\smash{v_\text{ND$_3^+$}}$ and $\lambda_\text{blue}$.
		(c) Relative intensity as a function of $E_\gamma$ and the final energy of the ion $E_f$ (relative to $\smash{X(N_K^p=1_1^+)}$). Horizontal black dashed lines mark the ionic vibrational thresholds as in \cref{fig:PI_spectra}. Relative intensity of (a,b) is expressed on a logarithmic color scale.
	}
	\label{fig:2D_spectra_11p}
\end{figure}

As such, this map also contains information on how Rydberg states populated at $E_\gamma$ autoionize to different ionic states with vibrational resolution. To show this more explicitly, \cref{fig:2D_spectra}(c) shows the same dataset with every column modified according to \cref{eq:Ef}. Note that the horizontal scale is the same throughout \cref{fig:2D_spectra}(a,b,c), and that the top axis can be converted into the bottom axis by converting to wavenumbers and adding the energy of the VUV photon. The three most intense horizontal bands in \cref{fig:2D_spectra}(c) can be assigned to $\nu_2^+=4,5,6$. The remaining two weaker traces lying between the $\nu_2^+$ bands show that another vibrational mode also plays a role in the dynamics. Based on a comparison of their energies and calculated vibrational term values for ND$_3^+$, these extra traces can likely be assigned to vibrational states with only a single quantum in $\nu_4$, the in-plane bending mode, and 3 or 4 quanta in the $\nu_2$ mode \cite{Leonard2002}. In general, the traces gradually decrease in intensity as $E_\gamma$ increases. When a new channel opens, the others show a sudden decrease in intensity. This is particularly visible in the $\nu_2^+=5$ trace when $\nu_2^+=6$ opens.

\Cref{fig:2D_spectra_11p} shows the same type of plots, this time following excitation through the $B(\nu'_2=5,N'_{K'}=2_2) \leftarrow X(\nu_2=0,N_K^p=1_1^+)$ transition. During scattering experiments, this excitation will be used to detect the $1_1^+$ collision products. Compared to \cref{fig:2D_spectra}, a narrower energy range is covered, zooming in around the strongest resonance. Furthermore, less intensity is observed in channels other than the main vibrational channel ($\nu_2^+=4$ in this case). Note that the relative color scale covers a five-fold larger range.

\begin{figure}[b!]
	\centering
	\includegraphics[scale=1.2]{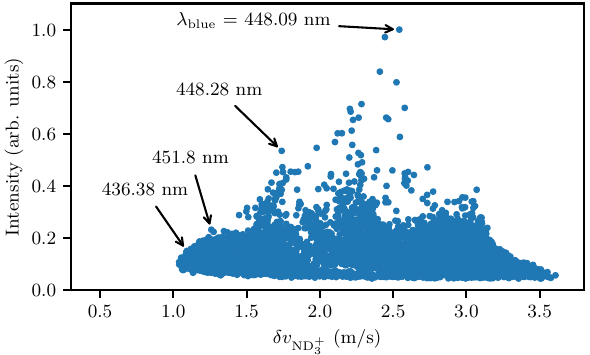}
	\caption{Scatter plot generated from the data set in \cref{fig:2D_spectra} by calculating the total intensity and resolution due to recoil $\smash{\delta v_\text{ND$_3^+$}}$ at each ionization wavelength $\lambda_\text{blue}$. ND$_3$ molecules were first excited through the $B(\nu'_2=6,N'_{K'}=2_2) \leftarrow X(\nu_2=0,N_K^p=1_1^-)$ transition.
	}
	\label{fig:recoil_vs_int}
\end{figure}

\begin{figure}[b!]
	\centering
	\includegraphics[scale=1.2]{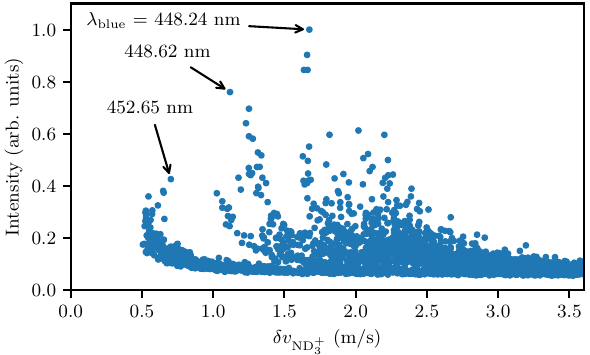}
	\caption{Scatter plot generated from the data set in \cref{fig:2D_spectra_11p} by calculating the total intensity and resolution due to recoil $\smash{\delta v_\text{ND$_3^+$}}$ at each ionization wavelength $\lambda_\text{blue}$. ND$_3$ molecules were first excited through the $B(\nu'_2=5,N'_{K'}=2_2) \leftarrow X(\nu_2=0,N_K^p=1_1^+)$ transition.
	}
	\label{fig:recoil_vs_int_11p}
\end{figure}

Unfortunately, \cref{fig:2D_spectra,fig:2D_spectra_11p} do not immediately reveal which blue wavelength offers the best trade-off between ion recoil and signal intensity. Therefore, we calculated and compared the resolution due to recoil $\smash{\delta v_\text{ND$_3^+$}}$ and summed intensity at every wavelength, as shown in \cref{fig:recoil_vs_int,fig:recoil_vs_int_11p} for detection of the $1_1^-$ and $1_1^+$ states, respectively. $\smash{\delta v_\text{ND$_3^+$}}$ is the `size' of the crushed and centered electron image (no Abel inversion), which reflects the distribution of ND$_3^+$ ions. But, this size is not well defined for images with multiple rings. We computed $\smash{\delta v_\text{ND$_3^+$}}$ as the mean radius weighted by the average intensity at each radius, yielding an estimate for the radius of the ion spotsize. This method underestimates the resolution of a step-like distribution (for which the fwhm is a more appropriate measure), but includes the contribution from a long tail or weak outer ring.

From the distribution of points in \cref{fig:recoil_vs_int} it is evident that the most intense transition of the spectrum at $\lambda_\text{blue}= \SI{448.09}{nm}$ is the optimal choice to detect $1_1^-$ for most experiments. It corresponds to a resolution of \SI{2.5}{m/s}, which greatly improves over the \SI{17}{m/s} of the 2+1 REMPI scheme. It performs similarly to other transitions with a relative intensity above $0.6$, so significantly lower recoil is only possible at the expense of a significant amount of intensity. One example is the transition at $\lambda_\text{blue}= \SI{448.28}{nm}$ ($0.53$ relative intensity, \SI{1.75}{m/s} recoil). The lowest recoil is obtained just above the $\nu^+_2=5,6$ thresholds, at $\lambda_\text{blue}= \SI{451.8}{nm}$ ($0.23$ relative intensity, \SI{1.26}{m/s} recoil) and $\lambda_\text{blue}= \SI{436.38}{nm}$ ($0.15$ relative intensity, \SI{1.1}{m/s} recoil), respectively. It is noted that all datapoints exceed \SI{1}{m/s}. This is due to a combination of the non-zero amount of background events, as well as the finite spot size even \SI{0}{m/s} electron events have on the detector. As such, the lowest values for $\smash{\delta v_\text{ND$_3^+$}}$ are not accurate, but still useful in comparing the different ionization wavelengths.

\Cref{fig:recoil_vs_int_11p} characterizes the transitions for detection of $1_1^+$ via the $\nu_2'=5$ intermediate. In general, the obtained recoil values are lower compared to \cref{fig:recoil_vs_int}, consistent with the fact that less intensity is observed in channels other than the main vibrational channel ($\nu_2^+=4$ in this case) when comparing the two 2D maps. The strongest transition at $\lambda_\text{blue}= \SI{448.24}{nm}$ yields a resolution of \SI{1.68}{m/s}. Again, the lowest recoil is obtained just above the $\nu^+_2=4$ threshold at $\lambda_\text{blue}= \SI{452.65}{nm}$ ($0.42$ relative intensity, \SI{0.70}{m/s} recoil). The transition at $\lambda_\text{blue}= \SI{448.62}{nm}$ ($0.76$ relative intensity, \SI{1.12}{m/s} recoil) offers a compromise between the two, in both intensity and resolution. All three transitions are good options to detect ND$_3$ ($1_1^+$) with near-zero recoil. For cold molecule scattering, we will use the most intense transition at $\lambda_\text{blue}=\SI{448.24}{nm}$ to obtain as much signal as possible. The true resolution of this 1+1$'$ REMPI detection scheme is demonstrated below by recording a scattering image.

\subsection{Rotationally Resolved Photoelectron Images}

\begin{figure}[b!]
	\centering
	\includegraphics{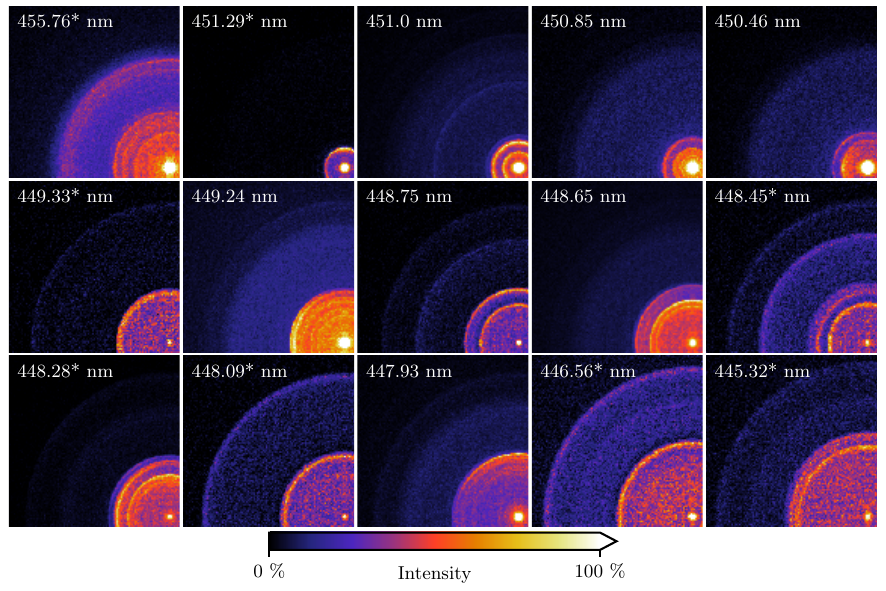}
	\caption{Rotationally resolved photoelectron images of ND$_3$ at several ionization wavelengths $\lambda_\text{blue}$, following $B(\nu'_2=6,N'_{K'}=2_2) \leftarrow X(\nu_2=0,N_K^p=1_1^-)$ excitation. Wavelengths followed by an asterisk are resonant with one of the observed autoionizing resonances. The images have been symmetrized, so only a single quandrant is shown. See the Experimental Section for further details on the image analysis. In addition, the images have been $2\times2$ rebinned for this figure.
	}
	\label{fig:PE_images}
\end{figure}

\begin{figure}[p!]
	\centering
	\includegraphics[scale=1.1]{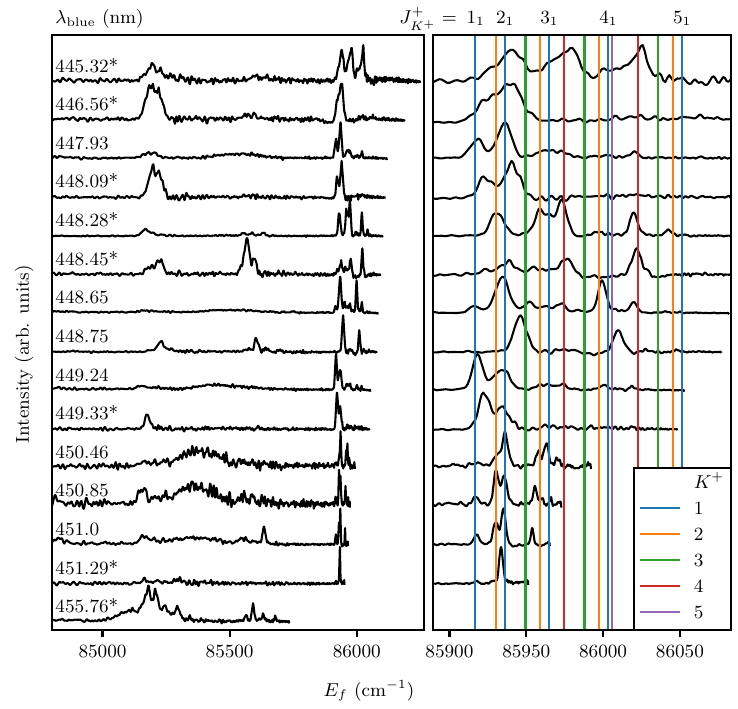}
	\caption{Rotationally resolved PE spectra of ND$_3$ at several ionization wavelengths $\lambda_\text{blue}$, following $B(\nu'_2=6,N'_{K'}=2_2) \leftarrow X(\nu_2=0,N_K^p=1_1^-)$ excitation. Wavelengths followed by an asterisk are resonant with one of the observed autoionizing resonances. The right panel shows a zoom-in of the $\nu_2^+=5$ threshold region (\num{85890} to \SI{86080}{\per\cm}), including the rotational assignment using the parameters from \cref{tab:rotational_constants}. Rotational lines are color coded by $K^+$, with the $K^+=1$ lines labelled along the top of the figure. Higher values of $K^+$ appear at lower energies. These spectra were obtained from the eVMI images in \cref{fig:PE_images} using a calibration factor of $f_\text{VMI}=1/120$~\si{cm^{-1}/pixel^2}. The intensities were multiplied by $R\propto\sqrt{K_e}$ to make the features at low $E_f$ more prominent.
	}
	\label{fig:Ke_rotational}
\end{figure}

Next, \cref{fig:PE_images} shows several high resolution photoelectron images that we recorded following excitation through the $B(\nu'_2=6,N'_{K'}=2_2) \leftarrow X(\nu_2=0,N_K^p=1_1^-)$ transition at different ionization wavelengths in an effort to resolve and assign the rotational structure of the $X^+(\nu_2^+=5)$ state, analogous to Ref.~\citenum{Hockett2010}. The resulting PE spectra are shown in \cref{fig:Ke_rotational}. Essentially, these represent columns in \cref{fig:2D_spectra}(c), albeit at a higher energy resolution. All spectra are plotted as a function of $E_f$ according to \cref{eq:Ef0}. The energy resolution improves at lower electron energy (higher $E_f$) in a given spectrum because the velocity resolution is constant when using VMI to record PE spectra. We primarily picked ionization wavelengths just above the $\nu_2^+=5$ threshold to achieve the best possible resolution. The assigned $X^+(\nu_2^+=5)$ rotational states are shown as vertical colored lines. Their position is given by the rotational parameters summarized in \cref{tab:rotational_constants}, whose values closely match those reported for the $F'$ state \cite{Langford1998}. Since the $F'$ state already has a high principal quantum number, its rotational parameters should closely match those of the ionic ground state.

\begin{table}[p!]
	\caption{Rotational Constants of ND$_3$ $X^+(\nu_2^+=5)$ Used to Assign the PE Spectra in \cref{fig:Ke_rotational} \textsuperscript{\emph{a}}}
	\label{tab:rotational_constants}
	\begin{tabular}{*4{c}}
		\hline \hline
		\noalign{\vspace{0.1cm}}
		$\nu_2^+$ & $E_0^+$ (\si{\per\cm}) & $B^+$ (\si{\per\cm}) & $C^+$ (\si{\per\cm}) \\
		\hline
		\noalign{\vspace{0.1cm}}
		5 & 85909 & 4.8 & 2.9 \\
		\hline \hline
	\end{tabular}
	\\ \vspace{0.2cm}
	\small{
		\textsuperscript{\emph{a}} $E_0^+$ is relative to $X(1_1^-)$ and has been corrected for the presence of the VMI field by \cref{eq:Ef0}.	
	}
\end{table}

A strong wavelength dependence for the final state distribution of the formed ion is observed again, now at the rotational level. This complicates an internal confirmation of the calibration factor and the PE peak assignments. Still, by recording images at closely spaced intervals of \SI{\sim0.2}{nm}, certain lines persistently appear across several PE spectra, restricting the assignment. Coincidentally, the top and bottom PE spectra show a similar, partly resolved rotational structure around the $\nu_2^+=4$ threshold at \SI{85150}{\per\cm}. Overlapping these features fixes the energy calibration factor to $1/120$~\si{cm^{-1}/pixel^2} with a high sensitivity, since these two spectra differ most in $\lambda_\text{blue}$. Now, the $\nu_2^+=4,5$ regions show a vibrational spacing of roughly \SI{745}{\per\cm}, again matching the $F'$ state \cite{Langford1998}. The bottom PE spectrum at $\lambda_\text{blue}=\SI{455.76}{nm}$ was recorded to resolve the rotational lines of the vibrational interloper just above \SI{85500}{\per\cm}. The four resolved features neatly overlap with the ones observed for $\lambda_\text{blue}=$\numlist{448.45;448.75;451.0}~\si{nm}, but the resolution is insufficient to make a full assignment. We assign the broad feature predominantly visible for $\lambda_\text{blue}=$\numlist{450.46;450.85}~\si{nm} as background electrons, since the feature appears at a constant energy relative to the onset of each PE spectrum. These background signals are also present at a lower intensity in the spectra for $\lambda_\text{blue}=$\numlist{455.76;451.00;449.24;447.93}~\si{nm}. The relative intensity of this background feature differs per image as the laser powers were reduced for each image separately to reach signal levels compatible with event counting.

Focusing on the $\nu_2^+=5$ threshold in \cref{fig:Ke_rotational}, the four spectra between $\lambda_\text{blue}=$\numlist{450.46;451.29}~\si{nm} resolve the lowest rotational states best, in particular the states $N^+_{K^+}=2_1$ and $2_2$. States with $K^+=1,2$ are visible across the PE spectra.
In general, states with $K^+=3$ are either absent or weak since they belong to the different $A_{1,2}$ nuclear spin symmetry. The assignment is not always perfect, due possibly to varying stray fields in the apparatus. For example, the strongest peak in the $\lambda_\text{blue}=\SI{451.29}{nm}$ spectrum falls in between the $2_1$ and $2_2$ lines, while a weak signal corresponding to $1_1$ is visible at the correct energy. Similarly, a peak is present in between the $3_2$ and $3_3$ lines for $\lambda_\text{blue}=$\numlist{451.00;450.85}~\si{nm}.

\subsection{Assignment of the Rydberg Spectra}
\label{subsec:MQDT}
To assign individual lines in the PI spectra one can search for the presence of Rydberg series spanning the spectra. To first approximation, the energy level of the $n$\textsuperscript{th} Rydberg state in a series converging to $X^+(\nu^+, N^+_{K^+})$ with adiabatic ionization energy $T_0(\nu^+, N^+_{K^+})$ is given by

\begin{equation}
	\label{eq:E_QD}
	E_n(\nu^+,N^+_{K^+},\ell_e,J) = T_0(\nu^+,N^+_{K^+}) - \frac{Rhc}{(n-\mu(\nu^+,N^+_{K^+},\ell_e,J))^2} \ ,
\end{equation}

\noindent where $R$ is the Rydberg constant, $\ell_e$ is the orbital angular momentum of the Rydberg electron, $J$ is the total angular momentum of the Rydberg electron and ion core, and $\mu(\nu^+,N^+_{K^+},\ell_e,J)$ is the quantum defect. In this work, as in Ref.~\citenum{Dickinson:JPCA105:5590}, we label the Rydberg states with the quantum number $J$, but assume that only pure singlet Rydberg series are accessed (i.e. the net spin for ion core and Rydberg electron is zero) and hence the $J$ quantum number is restricted to values derived from the vector sum of $N^+$ and $\ell_e$. The quantum defect is the only fit parameter for an unperturbed Rydberg series when the ionic energy levels are known. It is assumed to be energy-independent over limited ranges of the spectrum. Multiple Rydberg series can converge onto the same ionic state, differing by the orbital quantum numbers of the electron. A Rydberg series with the electron in a $p$-type orbital is also called an $np$ series. 

Symmetry considerations allow determination of the selection rules for which Rydberg series should be observable for molecules characterized in the $D_{3h}$ group. As discussed in Ref.~\citenum{Dickinson:JPCA105:5590}, the total symmetry of the final state of a transition (whether a Rydberg excitation or excitation to a continuum) is given by a product of symmetries of the ion core electronic state ($A_2''$), the ion core vibration ($A_1'$  for even, $A_2''$ for odd number of quanta in the $\nu_2$ mode), the rotational state of the ion (symmetry determined by the $K^+$ quantum number and the Rydberg electron or ionized electron ($A_1'$ for $l$ even, $A_1''$ for $l$ odd)). When exciting via the $B(\nu'_2=5,N'_{K'}=2_2)$ level the rovibronic symmetry of that intermediate level is $E'$ and selection rules require transitions to final states of total rovibronic symmetry $E''$. When exciting via the $B(\nu'_2=6,N'_{K'}=2_2)$ level the intermediate symmetry is $E''$ and the final state must be $E'$. The similarity in appearance of the spectra recorded via $\nu_2'=5$ and $6$ occurs because for the transitions diagonal in $\nu_2$ the vibrational symmetry of the final state is different in the two cases, and this results in the rotational symmetries allowed for the final state being the same. Following the considerations elaborated in Ref.~\citenum{Dickinson:JPCA105:5590} we predict that for $nd$ series excitation diagonal in vibrational quantum number (or with an even numbered change of $\nu_2$), the core rotational quantum numbers accessible are principally the $E''$ series, $N^+ = 1-5$, $K^+=1$, or $N^+=5$, $K^+=5$ while for $np$ series it is $N^+=2-4$, $K^+=2$ and $N^+=4$, $K^+=4$. 
For odd numbered changes in $\nu_2$, the series expected are: for $nd$ excitation $N^+ =2-5$, $K^+=2$ or $N^+=4-5$, $K^+=4$; for $np$ excitation $N^+=2-4$, $K^+=1$. The intermediate $B$ state level of the transitions has quantum numbers $J'=2$, $K'=2$, hence the final angular momentum must be constrained to $J = 1,2$ or $3$. In general we expect negligible differences in quantum defects for different $J$ values for the $nd$ series, whereas significant differences of $\mu$ may be found for the $np$ series of different $J$. Finally, given that the $B$ state is primarily of $np$ character but with a contribution of $nd$ character (see Ref.~\citenum{Dickinson:JPCA105:5590}) we could in principle observe $ns$ and $nf$ series.  However we did not find any evidence for series with those quantum defects in the spectra and in previous simulations of the MATI spectra it was not necessary to include $nf$ excitations \cite{Dickinson:JPCA105:5590}.

\begin{table}[p!]
	\caption{Parameters Used to Assign the Rydberg Series of ND$_3$	\textsuperscript{\emph{a}}} 
	\label{tab:Rydberg_parameters_TSoftley}
	\begin{tabularx}{\textwidth}{*4{X}}
		\hline \hline
		\noalign{\vspace{0.1cm}}
		{\raggedright vibrational level/ adiabatic limit ($N^+=0$, $K^+=0$) from ground state level (\si{\per\cm})} & {\raggedright ionic state of convergence/ Rydberg symmetry} & {\raggedright quantum defect $[J]$} & {\raggedright wavenumber from ground state level ($1_1$) to the zero-field threshold (\si{\per\cm})}  \\
		\hline
		\noalign{\vspace{0.1cm}}
		$\nu_2^+=6$ &   $N^+=1$, $K^+=1$, $nd$  &   0.033   &   \num{86684.6} \\
		&	$N^+=2$, $K^+=1$, $nd$	&	0.032	&	\num{86703.4}	\\
		\num{86677.1}	&	$N^+=3$, $K^+=1$, $nd$	&	0.035	&	\num{86731.6}	\\
		&	$N^+=4$, $K^+=1$, $nd$	&	0.028	&	\num{86769.2}	\\
		&	$N^+=5$, $K^+=1$, $nd$	&	0.025	&	\num{86816.2}	\\
		&	$N^+=5$, $K^+=5$, $nd$	&	0.042	&	\num{86770.6}	\\
		&		&		&		\\
		&	$N^+=2$, $K^+=2$, $np$	&	0.657 [3]	&	\num{86697.7}	\\
		&							&	0.501 [2]	&		\\
		&							&	0.744 [1]	&		\\
		&	$N^+=3$, $K^+=2$, $np$	&	0.623	&	\num{86725.9}	\\
		&	$N^+=4$, $K^+=2$, $np$	&	0.588	&	\num{86763.5}	\\
		&	$N^+=4$, $K^+=4$, $np$	&	0.744	&	\num{86740.7}	\\
		&		&		&		\\
		&		&		&		\\
		$\nu_2^+=5$	&	$N^+=2$, $K^+=2$, $nd$	&	0.027	&	\num{85931.7}	\\
		&	$N^+=3$, $K^+=2$, $nd$	&	0.022	&	\num{85959.9}	\\
		\num{85911.1}	&	$N^+=4$, $K^+=2$, $nd$	&	0.026	&	\num{85997.5}	\\
		&	$N^+=5$, $K^+=2$, $nd$	&	0.026	&	\num{86044.5}	\\
		&	$N^+=4$, $K^+=4$, $nd$	&	0.03	&	\num{85974.7}	\\
		&	$N^+=5$, $K^+=4$, $nd$	&	0.034	&	\num{86021.7}	\\
		&		&		&		\\
		&	$N^+=1$, $K^+=1$, $np$	&	0.592 [2]  	&	\num{85918.6}	\\
		&							&	0.372 [1]	&		\\
		&	$N^+=2$, $K^+=1$, $np$	&	0.547 [3]  	&	\num{85937.4}	\\
		&							&	0.492 [2]	&		\\
		&							&	0.372 [1]	&		\\
		&	$N^+=3$, $K^+=1$, $np$	&	0.522 [3]  	&	\num{85965.6}	\\
		&							&	0.397 [2]	&		\\
		&	$N^+=4$, $K^+=1$, $np$	&	0.412	&	\num{86003.2}	\\
		
		\hline \hline
	\end{tabularx}
	\\ \vspace{0.2cm}
	\small{
		\textsuperscript{\emph{a}} The rotational constants used to calculate the thresholds were $B_5=4.9$, $C_5=2.8$, $B_6=4.9$ and $C_6=2.8$. The quantum defects were taken from the diagonal elements of the MQDT matrix in the Hund’s case d representation (see Ref.~\citenum{Duggan2010}, Eq.~14). All $nd$ series are assumed to have quantum defects independent of the $J$ quantum number (hence no $J$ splittings are observed), although the transitions to $J=3$ are expected to be strongest.  For the $np$ series the $J$ quantum number is indicated in square brackets after the quantum defect where $J=1$ and $2$ series are distinctly observed.
	}
\end{table}

\begin{figure}
	\centering
	\begin{overpic}[trim={3cm 0 0 0},clip,width=0.95\linewidth]{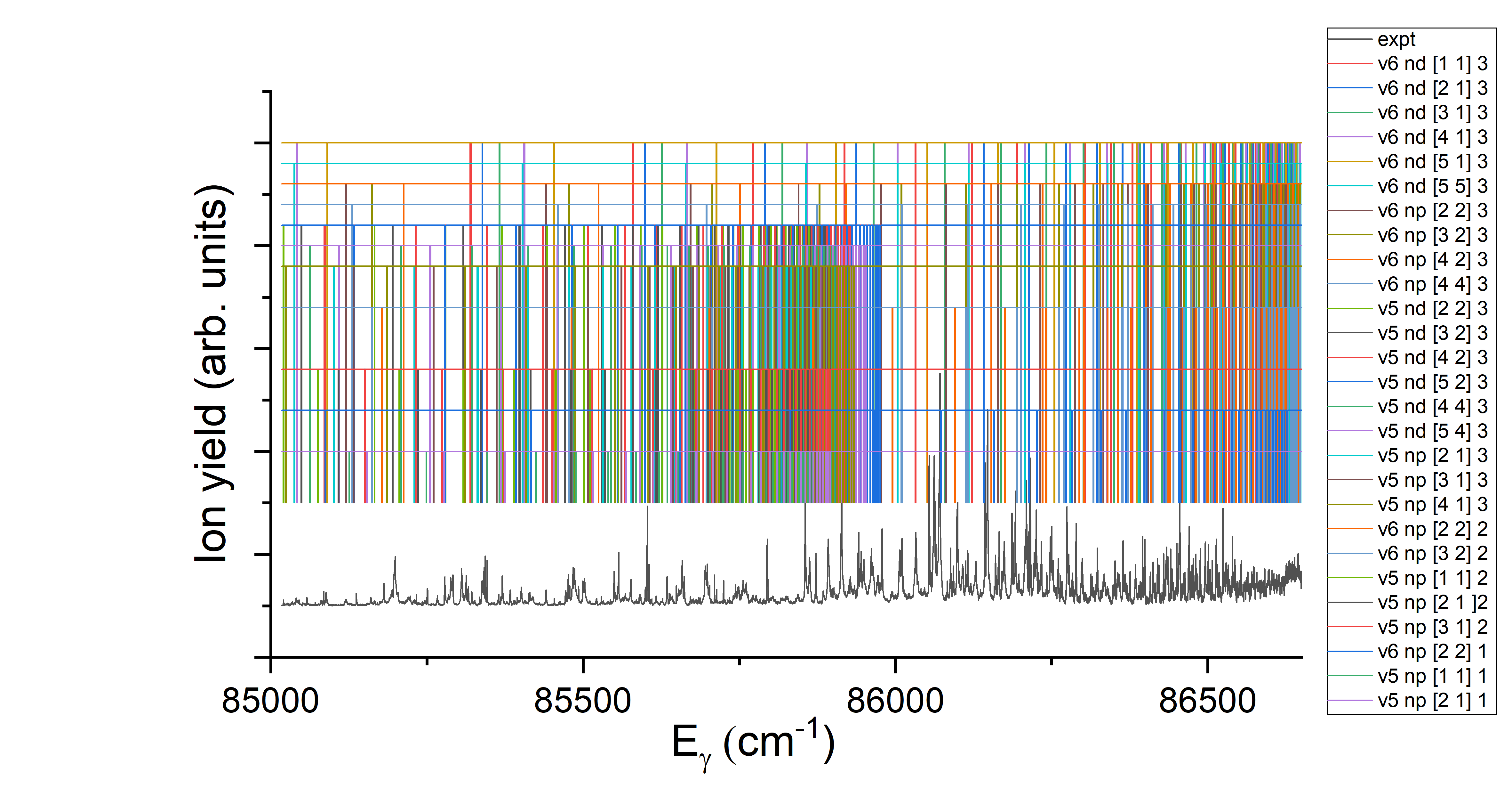}
		\put(0,45) {(a)}
	\end{overpic} \\
	\begin{overpic}[trim={3cm 0 0 0},clip,width=0.95\linewidth]{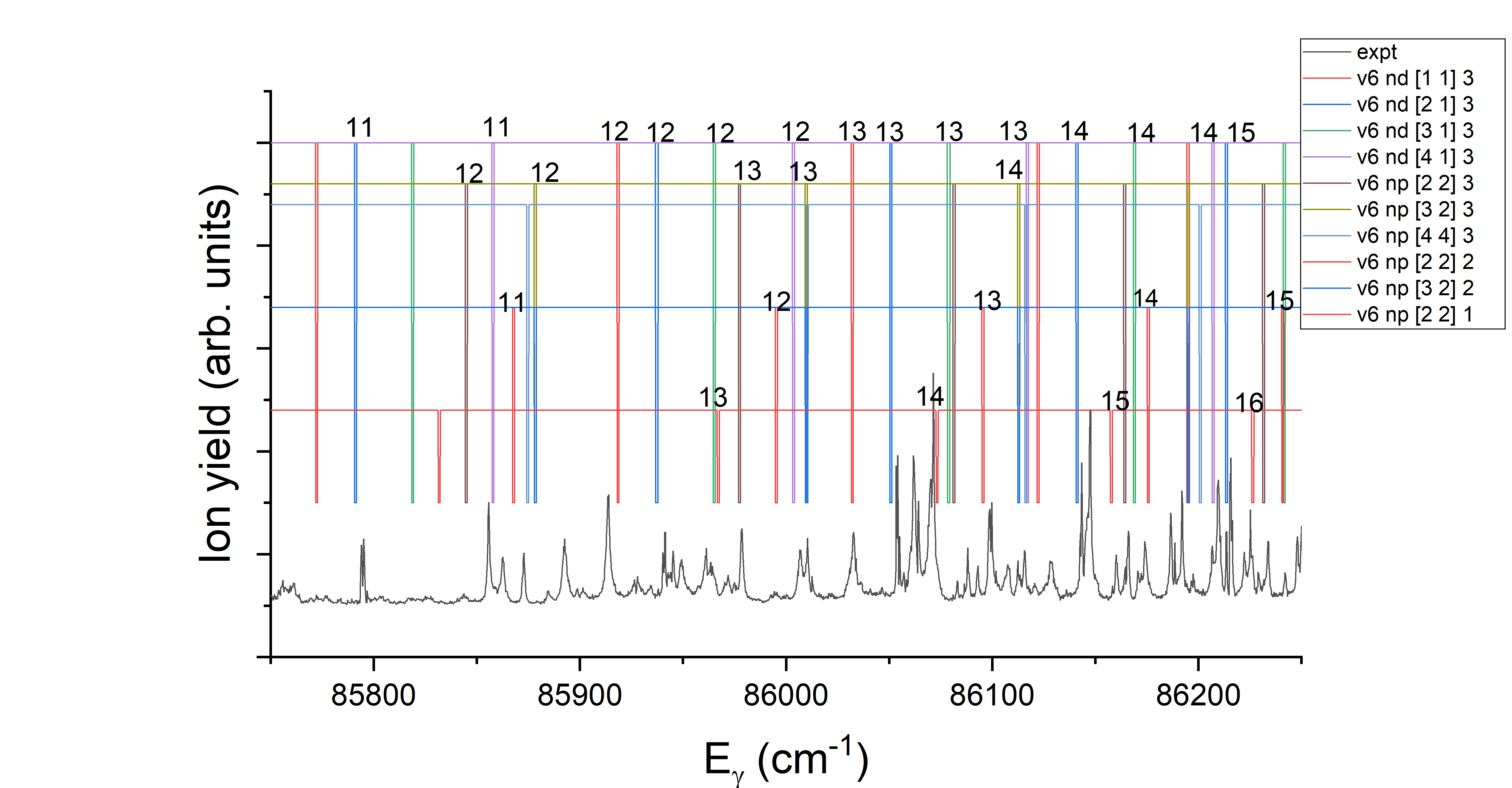}
		\put(0,45) {(b)}
	\end{overpic}
	\caption{Preliminary assignment of the PI spectrum of ND$_3$ displayed in \cref{fig:PI_spectra}(a), following excitation through $B(\nu'_2=6)$. (a) full spectrum, (b) zoomed-in region including the strongest observed resonance at $E_\gamma = \SI{86070}{\per\cm}$. The experimental PI spectrum is shown in black. Twenty-seven separate Rydberg series are displayed above the PI spectrum in (a), as discussed in the text, and calculated from the Rydberg formula (\cref{eq:E_QD}) with quantum defects as in \cref{tab:Rydberg_parameters_TSoftley}. In (b) a subset of the series are shown, and the principal quantum number $n$ is listed along the top. Each series is identified in the legend. For example, ``v6 nd [2,1]3'' signifies the $nd$ Rydberg series converging toward the $\nu_2^+=6$, $N^+_{K^+}=2_1$ ionic state, with total angular momentum $J=3$.
	}
	\label{fig:PI_Rydberg_figure_TSoftley}
\end{figure}

The assignment of the PE spectra as presented in the previous section determines $T_0(\nu_2^+=5,N^+_{K^+})$. Similarly, $T_0(\nu_2^+=6,N^+_{K^+})$ follows from the \SI{766.7}{\per\cm} vibrational spacing found by overlapping the two PI spectra shown above and assuming similar rotational parameters. Starting from these assumptions, and the diagonal quantum defect parameters reported in Refs.~\citenum{Dickinson:JPCA105:5590,Duggan2010} we currently identify 27 separate Rydberg series in a preliminary assignment of the PI spectrum following excitation through the $B(\nu'_2=6) \leftarrow X(N_K^p=1_1^-)$ transition. Small adjustments were made to the quantum defects to improve the fit and \cref{tab:Rydberg_parameters_TSoftley} lists all the parameters used in the Rydberg formula for these assignments. \Cref{fig:PI_Rydberg_figure_TSoftley}(a) shows these Rydberg series plotted above the PI spectrum. The most clearly defined Rydberg series include seven that converge onto $\nu_2^+=6$ (the $nd$ series with $N^+_{K^+}=1_1, 2_1, 3_1, 4_1$, and the $np$ series converging to $2_2$, $3_2$ and $4_4$), and (below \SI{86000}{\per\cm}) six that converge onto $\nu_2^+=5$ ($nd$ series to $N^+_{K^+}=2_2, 3_2, 4_2$, and $np$ series to $2_1, 3_1, 4_1$). \Cref{fig:PI_Rydberg_figure_TSoftley}(b) shows the same spectrum, zoomed-in around the strongest resonance at $E_\gamma = \SI{86070}{cm}$. Only a few of the $\nu_2^+=6$ Rydberg series are shown here, with the principal quantum numbers $n$ listed along the top. Most of the Rydberg series shown in \cref{fig:PI_Rydberg_figure_TSoftley}(b) are labelled as $J=3$, but three $np$ series are identified as $J=2$ or $1$, which have significantly different quantum defects (as calculated using Eq.~14 of Ref.~\citenum{Duggan2010}) to the $J=3$ series with the same $N^+$, $K^+$, $\ell_e$ values. The strongest resonance can be assigned to a Rydberg state with principal quantum number $n=14$ of an $J=1$ $np$ series with $N^+_{K^+}=2_2$.

With this approach, a large part of the spectrum can be tentatively assigned. However, it is only possible this way to assign line positions. No prediction can be made on the line strengths or widths. More accurate modeling of the PI spectra requires an MQDT simulation, as described in Refs.~\citenum{Ross1991,Dickinson:JPCA105:5590,Duggan2010}. This method can be used to describe both bound states and continua, which coexist in the threshold region studied here, and the interactions between them. Excitation to the Rydberg series and the associated photoionization continua are calculated by developing an electron---ion-core scattering wave function. The electron---ion-core interaction may cause a variety of effects including autoionization and perturbation of spectral line position, especially where members of two different Rydberg series come close in energy. This perturbation may also cause intensity and line width changes, and, in general, deviations from the simple Rydberg formula, \cref{eq:E_QD}. In MQDT, all interactions between the channels are encoded in the quantum defect matrix $\bm{\mu}$ representing the magnitude of interactions when the electron penetrates into the core region. In a forthcoming publication, we will describe the use of MQDT to model these spectra and the photoelectron spectra in detail.

\subsection{Scattering Image}
\label{subsec:VUV_recoil_comp}
\begin{figure}[b!]
	\centering
	\includegraphics[scale=1.2]{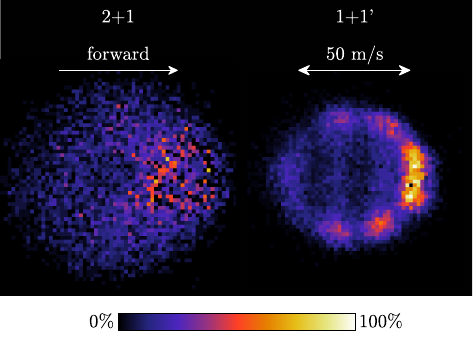}
	\caption{Scattering images for the inelastic scattering of ND$_3$ ($1_1^- \rightarrow 1_1^+$) with HD($j=0$) at a collision energy of \SI{5.7}{\per\cm}. The images are recorded with the \SI{17}{m/s} recoil 2+1 REMPI scheme at \SI{321}{nm} (left) and low-recoil 1+1$'$ REMPI scheme (right). For the latter, the strongest peak in the PI spectrum was used with wavelengths as in \cref{tab:laser_wavelengths_VUV}. This yielded maximum signal, with a recoil of only \SI{1.7}{m/s} (\cref{fig:recoil_vs_int_11p}).
	}
	\label{fig:DCS_recoil_comparison}
\end{figure}

Finally, we recorded two crossed molecular beam scattering images to compare the effect of ion recoil induced by the 2+1 and 1+1$'$ REMPI schemes directly, which are shown in \cref{fig:DCS_recoil_comparison}. Both images are obtained under the same experimental conditions, except for the ionization scheme. They show inelastic scattering of ND$_3$ ($1_1^- \rightarrow 1_1^+$) with HD($j=0$) at a collision energy of \SI{5.7}{\per\cm}.

In the scattering image recorded with the 2+1 REMPI scheme, only a preference to forward scattering can be discerned. The \SI{17}{m/s} recoil is comparable to the image radius itself, washing out all fine details and spoiling the high velocity resolution that is enabled by the use of a Stark decelerator. On the other hand, the scattering image recorded with the 1+1$'$ REMPI scheme has a resolution comparable to the velocity spread of the ND$_3$ beam (\SI{\sim5}{m/s}), and hence is not limited by recoil. This image resolves a weak backscattering feature, together with two sidescattering features, whose crushed contribution is visible as two bands crossing the center of the image. We found that the 2+1 and 1+1$'$ detection schemes lead to comparable signal levels. The low conversion efficiency during the DFM mixing process is compensated by the large one-photon absorption cross section in ammonia. This low-recoil 1+1$'$ REMPI scheme can be used in future experiments to image ammonia with a high velocity resolution.

\section{Conclusions}
We introduced a new low-recoil detection scheme for ND$_3$ by performing 1+1$'$ REMPI through the $B(\nu'_2=5,6)$ states. The required VUV photons of \SI{160}{nm} were generated through DFM in Xe, and ionization proceeded by further excitation into the $X^+(\nu^+_2=5,6)$ threshold region. We observed a strong propensity for excitation to Rydberg states below the diagonal ($\Delta\nu_2=0$) ionization threshold, which efficiently autoionized with little recoil. By velocity mapping the photoelectrons, it was possible to record wavelength dependent PE spectra, generating a 2D map of the photoionization dynamics with vibrational resolution and identifying the transitions best suited for efficient, low-recoil detection. The photoionization dynamics could be rotationally resolved in PE spectra from eVMI images. The PI spectra could be tentatively assigned by identifying multiple Rydberg series, using parameters from the PE measurements to restrict the model. Finally, we demonstrated the effectiveness of this low-recoil detection scheme for recording high resolution scattering images in cold collision experiments. The 1+1$'$ scheme features a recoil of only \SI{1.7}{m/s}, with a sensitivity comparable to the conventional 2+1 scheme which has a recoil of \SI{17}{m/s}. This paves the way for future scattering studies involving ND$_3$, and encourages the use of VUV radiation in the search for low-recoil REMPI schemes of other species.

\begin{acknowledgement}

The authors thank R. Louwerse and N. Schrijer for their adjacent work on the photoelectron spectroscopy of ammonia, and A. van Roij and N. Janssen for excellent technical support. T.P.S. is grateful to the Leverhulme Trust for financial support of his contribution to this work. S.Y.T.v.d.M. acknowledges support from the European Research Council (ERC) under the European Union’s Horizon 2020 Research and Innovation Program (Grant Agreement No. 817947 FICOMOL).

\end{acknowledgement}

%
%


\bibliography{bibliography}

\end{document}